\documentclass[aps,prd,nofootinbib,twocolumn,superscriptaddress,preprintnumbers]{revtex4}

\usepackage{amssymb}
\usepackage{amsmath}
\usepackage{epsfig}
\usepackage{hyperref}
\usepackage{breakurl,comment}

\newcommand{\be}{\begin{equation}}
\newcommand{\ee}{\end{equation}}
\newcommand{\bea}{\begin{eqnarray}}
\newcommand{\eea}{\end{eqnarray}}

\newlength {\leftpanesize}
\newlength {\rightpanesize}
\newlength {\titleskip}
\newlength {\startskip}
\newlength {\itemskip}
\newlength {\panesep}
\newlength {\itemparskip}
\newlength {\locationskip}
\newlength {\timeskip}
\newlength {\languagesize}

\begin{document}
\title{Multi-Vacuum Initial Conditions and the Arrow of Time}
\author{Raphael Bousso}
\author{Claire Zukowski}

\affiliation{Center for Theoretical Physics and Department of Physics, University of California, Berkeley, CA 94720, U.S.A}

\affiliation{Lawrence Berkeley National Laboratory, Berkeley, CA 94720, U.S.A}

\begin{abstract}
Depending on the type and arrangement of metastable vacua in the theory, initial conditions in a de~Sitter vacuum with arbitrarily large entropy can be compatible with the observed arrow of time, if the causal patch or related measures are used to regulate divergences.  An important condition, however, is that the initial vacuum cannot produce observers from rare fluctuations (Boltzmann brains).  Here we consider more general initial conditions where multiple vacua have nonzero initial probability.  We examine whether the prediction of an arrow of time is destroyed by a small initial admixture of vacua that can produce Boltzmann brains.  We identify general criteria and apply them to two nontrivial examples of such initial probability distributions.  The Hartle-Hawking state is superexponentially dominated by the vacuum with smallest positive cosmological constant, so one might expect that other initial vacua can be neglected; but in fact, their inclusion drastically narrows the range of theory parameters for which an arrow of time is predicted.  The dominant eigenvector of the global rate equation of eternal inflation is dominated by the longest-lived metastable vacuum.  If an arrow of time emerges in the single-initial-vacuum approximation, then we find that this conclusion survives the admixture of other initial vacua.  By global-local measure duality, this result amounts to a successful consistency test of certain global cutoffs, including light-cone time and scale-factor time.
\end{abstract}

\maketitle

\section{Introduction}

The entropy in our past light-cone has increased by $\Delta S > 10^{103}$ since the era of big-bang nucleosynthesis.  Moreover, the universe is still far from a state of maximum entropy today.  The arrow-of-time problem is the challenge of explaining these observations.

Naively, it seems sufficient to posit an initial state of low entropy, at the earliest time when a semi-classical description of the universe is possible.  However, it has recently become clear that this assumption is neither necessary nor sufficient for the observed arrow of time.   The vacuum structure of the underlying theory plays a crucial part.  This is particularly important if the theory contains vacua with positive cosmological constant, as it must~\cite{Perlmutter:1998np,Riess:1998cb}.

In a stable de~Sitter vacuum, no arrow of time is predicted, independently of initial conditions~\cite{Dyson:2002pf}.  Conversely, there are vacuum landscapes such as the string landscape, which lead to an arrow of time even if the initial entropy is arbitrarily larger than the entropy at the time of nucleosynthesis~\cite{Bousso:2011aa}. This result builds on earlier analyses of the possible predominance of Boltzmann brains~\cite{Dyson:2002pf, Bousso:2008hz, DeSimone:2008if}. A Boltzmann brain is an observer produced by a minimal fluctuation from equilibrium.  Such observers see only an arrow of time large enough for their own existence.  They do not see a highly ordered world around them. The dominance of Boltzmann brains is equivalent to the absence of an arrow of time.

All previous analyses~\cite{Dyson:2002pf, Bousso:2008hz, DeSimone:2008if,Bousso:2011aa} relied on the simplifying assumption that initial conditions have support only in a single vacuum, in identifying conditions for the dominance or absence of Boltzmann brains. In general, however, a theory of initial conditions may assign nonzero probability to a number of different vacua. Initial conditions may even have some support in excited states above the metastable vacua. In this paper, we will explore how generalized initial conditions affect the conditions under which an arrow of time emerges.

There is a reason to suspect that a small correction to single-vacuum initial conditions might destroy the prediction of an arrow of time, even if nearly all of the initial probability is concentrated in one vacuum.   A necessary condition for an arrow of time is the {\em complete\/} inability of the initial vacuum to produce Boltzmann brains directly.  This is not a problem in a realistic landscape, since one expects that the overwhelming majority of vacua, including the one selected by initial conditions, will not give rise to the fine-tuned low-energy physics that allows for complex structures such as observers to exist, whether they form by classical evolution or by fluctuations.  However, if all or most vacua have at least a tiny probability to be the initial state, then this will include fine-tuned vacua which can produce Boltzmann brains, such as our own vacuum.  A more careful quantitative analysis is needed to understand how much initial probability in Boltzmann-producing vacua can be tolerated such that Boltzmann brains remain nevertheless suppressed.

A realistic vacuum structure must contain our own vacuum.  This means it must contain at least one de~Sitter vacuum that is stable over cosmological timescales.  Moreover, observation tells us that the age of the universe is comparable to the timescale set by the cosmological constant.  This implies that the universe is eternally inflating unless the decay rate of our vacuum is fine-tuned~\cite{Guth:1982pn}.  Eternal inflation requires a regulator, or measure.  Here, we use the causal patch measure~\cite{Bousso:2006ev, Bousso:2006ge}.  Closely related local measures, such as the fat geodesic~\cite{Bousso:2008hz, Larsen:2011mi} or the Hubbletube~\cite{Bousso:2012cv} are equivalent for the purpose of the (relatively crude) question of whether an arrow of time is predicted. Finally, there also exist global measures, such as light-cone time~\cite{Bousso:2009dm, Bousso:2009mw, Bousso:2010id} and scale factor time~\cite{DeSimone:2008bq}. They are exactly equivalent to local measures with a particular choice of initial conditions. (Indeed, the initial conditions we study in Sec.~\ref{dominanteigenvector} are those dictated by this equivalence. They are dominated by the longest-lived vacuum, and we determine whether they can be approximated by it for the purpose of the arrow of time.)

We assume that singularities are terminal (see Ref.~\cite{Garriga:2012bc} for an alternative assumption).  Worldlines that enter vacua with negative cosmological constant end at the big crunch. The importance of terminal vacua for the existence of {\em some\/} arrow of time was recently emphasized in Refs.~\cite{Harlow:2011az, Susskind:2012pp}.  In this paper, our interest is in a stronger condition required for agreement with observation: that the flow towards terminal vacua is strong enough to avoid the dominance of Boltzmann brains over ordinary observers.\looseness=-1

\emph{Outline and Summary}: We establish notation and briefly review important approximation techniques in Sec.~\ref{sec-notation}.  In Sec.~\ref{toy}, we gain some intuition by considering two toy landscapes; for each, we study conditions for the generalization from single- to multiple-vacuum initial conditions to destroy the prediction of an arrow of time.  

In Sec.~\ref{sec-large}, we consider a general landscape subject to certain assumptions about its structure; the string landscape is expected to satisfy these assumptions.  We briefly review the analysis of the arrow of time in the case of single-vacuum initial conditions~\cite{Bousso:2011aa}.  We then consider multiple-vacuum initial conditions.  We identify sufficient conditions both for the absence, and for the presence of an arrow of time.  

In the remaining sections, we consider two specific proposals for multiple-vacuum initial conditions that may give large weight to vacua with small cosmological constant.  (This is the only class of proposals of for which the question is nontrivial.  With initial conditions in vacua with large cosmological constant---e.g., Refs.~\cite{Lin84b, PhysRevD.27.2848}---an arrow of time is typically predicted.)  

In Sec.~\ref{HH}, we show that Hartle-Hawking initial conditions lead to Boltzmann brain dominance unless the initial vacuum that dominates paths to ordinary observers is extremely long-lived. This is a significant deviation from the result that would be obtained in the single-initial-vacuum approximation, even though the initial probability distribution is overwhelmingly dominated by a single vacuum.\footnote{We thank Don Page for a discussion that brought this possibility to our attention.}

In Sec.~\ref{dominanteigenvector}, we consider initial conditions set by the dominant eigenvector of the rate equations of eternal inflation.  These initial conditions are selected by global duals of the local measures we consider; they describe the late-time attractor regime of the global solution.  The distribution is dominated by the longest-lived metastable vacuum, and we find that the single-initial-vacuum approximation is reliable in this case.  We show that the probabilities for other initial vacua are much smaller than the amplitude to transition dynamically to such vacua from the longest-lived vacuum.  Thus, if an arrow of time is predicted in the single-initial-vacuum approximation, the same conclusion is obtained with the full eigenvector.   This implies that global measures such as the light-cone time cutoff or scale factor time cutoff are in accord with the observed arrow of time subject to certain conditions on the vacuum structure.

\section{Conventions and Approximations}
\label{sec-notation}

\subsection{Branching Ratios}
A landscape has a collection of stable or metastable vacua labeled by the index $i$, each with a cosmological constant $\Lambda_i$. For simplicity we assume these vacua have been labeled in order of increasing cosmological constant,
\be \Lambda_1 < \Lambda_2 < ... < \Lambda_N~. \ee
When $\Lambda_i$ is negative, the vacuum $i$ is terminal, in the sense that a geodesic entering a pocket universe with $\Lambda<0$ will terminate on a future singularity, the big crunch.

Any vacuum $j$ has a decay rate per unit four-volume $\Gamma_{ij}$ (which could be extremely small) to a different vacuum $i$. The total decay rate of vacuum $j$ is defined as the sum of all possible decay rates from $j$,
\be \Gamma_j = \sum_i \Gamma_{ij}~. \ee 

It is often convenient to define a dimensionless decay rate $\kappa_{ij}$,
\begin{align}
\kappa_{ij} = \frac{4\pi \Gamma_{ij}}{3H_j^4}~,
\end{align}
where
\be H_i = \left(\frac{\Lambda_i}{3}\right)^{1/2}\ee
is the expansion rate for non-terminal vacua at late times. The total dimensionless decay rate is defined as
\be \kappa_j = \sum_i \kappa_{ij}~. \ee

The branching ratio from vacuum $j$ to $i$ is defined as the corresponding decay rate divided by the total decay rate of the parent vacuum $j$,
\be \beta_{ij} = \frac{\Gamma_{ij}}{\Gamma_j} = \frac{\kappa_{ij}}{\kappa_j}~. \ee

At a coarse-grained level, the expected number of times a geodesic starting in vacuum $j$ enters vacuum $i$ is given by a sum over products of branching ratios along all possible paths from $j$ to $i$~\cite{Bousso:2006ev},
\be e_{ij} = \sum_p \sum_{i_1,i_2,...,i_{p-1}} \beta_{ii_{p-1}}\cdots \beta_{i_1 j}~. \label{geodesics2}\ee
The sum over $p$ indicates a sum over paths of any length.

\subsection{Double Exponential Arithmetic}
In this section we review the arithmetic of double exponential numbers. An exponentially large (or small) number is one of the form $e^x$ (or $e^{-x}$), where $x$ is large. We denote this by a double inequality, $e^x \gg 1$. A double exponentially large (or small) number is one of the form $e^x$ (or $e^{-x}$) with $x\gg 1$, which we denote by a triple inequality $e^x \ggg 1$. For example, the Boltzmann brain production rate $\Gamma_{BB} < e^{-S_{BB}}$, where $S_{BB}$ is the minimum entropy required to create a Boltzmann brain, is a double exponentially small number. (The inequality here comes from the fact that a Boltzmann brain requires a minimum free energy of the order $S_{BB}$, so its production is accordingly suppressed.)

If $x$ and $y$ are at least exponentially large and $x>y$, then
\be x\pm y \approx x~. \ee

Assuming $x$ and $y$ are \emph{double} exponentially large, we can apply the previous rule to the exponent to obtain a product rule. Again assuming $x>y$,
\be xy \approx \frac{x}{y} \approx x~. \ee
Similar rules apply for exponentially or double exponentially small numbers, but with $x$ replaced by $1/x$. 

We can apply this arithmetic to landscape decay rates to obtain useful approximate identities. For example,

\be \frac{\Gamma_{BB,i}}{\Gamma_i} \approx  \begin{cases}
   \Gamma_{BB,i} < e^{-S_{BB}}   & \text{if } \Gamma_{BB,i} < \Gamma_{i}~,\\
   \Gamma_{i}^{-1} \ > e^{S_{BB}} & \text{if } \Gamma_{BB,i} > \Gamma_{i}~, \end{cases} \label{ratio1}\ee
where $\Gamma_{BB,i}$ is the Boltzmann brain production rate in vacuum $i$.

\subsection{Dominant History Method}
In an eternally inflating space-time, the number of occurrences of different types are infinite. This means that the relative probability of event $I$ compared to event $J$,
\be \frac{P_I}{P_J} = \frac{\left< N_I\right>}{\left< N_J\right>}~, \ee
where $\left<N_I\right>$ is the expected number of events of type $I$, which is infinite, is ill-defined. To address this problem, we must introduce a cutoff procedure that regulates infinities by counting only a finite subset of the events $I$. 

In this paper, we use the causal patch cutoff~\cite{Bousso:2006ge}. Probabilities are defined by counting events in a single causally connected region of space-time (the ``patch''), in a weighted average over initial conditions and decoherent histories of the patch. In a theory with long-lived vacua $i$, the expected number of events $\left< N_I\right>$ of type $I$ can be computed as
\be \left< N_I\right> = \sum_{i,j} N_{Ii}  e_{ij} P_j~. \label{expectedn}\ee
Here $P_j$ is the probability of starting in vacuum $j$; $e_{ij}$ is the probability that a geodesic that starts in $j$ will enter vacuum $i$,  given by Eq.~(\ref{geodesics2}); and $N_{Ii}$ is the number of events of type $I$ in vacuum $i$, within the patch.

Here the events of interest, $I$, will be observations made by Boltzmann brains, $BB$s, versus those made by ordinary observers, $OO$s. The expected number of $BB$s produced in vacuum $i$ is equal to the lifetime of vacuum $i$ multiplied by the rate of $BB$ production in $i$, so $N_{BB, i} = \Gamma_{BB, i}/\Gamma_i$. In the case $\Gamma_{BB,i} < \Gamma_i$, where $\Gamma_{BB,i}$ is a double-exponentially small number, this can be represented by a branching ratio, if decay channels are augmented by a ``decay to Boltzmann brains'':
\be N_{BB,i} \approx \beta_{BB,i} \equiv \frac{\Gamma_{BB,i}}{\Gamma_i + \Gamma_{BB,i}}~, \label{newbranchingratios}\ee
By double exponential arithmetic, Eq.~(\ref{ratio1}), we can also approximate $N_{BB, i}\approx \Gamma_{BB,i}$.
Numbers that are not double-exponentials, such as the precise number of ordinary observers within a patch that contains any, and the number of observations made by each, can be set to unity.\looseness=-1

There is a convenient method for estimating probabilities using the causal patch cutoff, based on its branching-tree implementation \cite{Bousso:2006ev, Bousso:2007er}.  A path through the landscape is represented as a sequence of arrows between vacua along the path. The branching ratios for each individual process are written above these arrows, using the notation $1'$ to represent a branching ratio that is one up to a small correction. 

In addition to labels denoting vacua (which denote generic states in these vacua, e.g., empty de~Sitter space), we also include labels ``$OO$" and ``$BB$" to represent ordinary observers and Boltzmann brains. Whether such observers form depends not only on the vacuum, but also on how it is approached along the decay path. If ordinary observers are produced by normal dynamical evolution after the decay of a higher $\Lambda$ vacuum, then we denote the decay as branching to $OO$, with the branching ratio set by the decay rate of the parent vacuum.  This is followed by an eventual decay to empty de~Sitter space, with branching ratio $1'$.  For Boltzmann brains, we use Eq.~(\ref{newbranchingratios}).

However, if no observers are produced in the approach to equilibrium in the new vacuum, then ordinary observer production requires an extra up-tunneling.  It is then even more suppressed than Boltzmann brains, since the latter require a much smaller entropy decrease: $\Gamma_{OO} \lll \Gamma_{BB} < e^{-S_{BB}}$.  This is denoted by decay to a de~Sitter vacuum followed by another, highly suppressed ``decay'' to observers.

To account for general initial conditions, each path begins with the label ``$I.C.$" followed by an arrow to some vacuum. The probability of starting in this vacuum will be written above the arrow.  

The total amplitude for ordinary observers or Boltzmann brains is obtained by summing over all paths that include such observers.  In realistic models, most branching ratios and initial probabilities will be double-exponentially small, and each sum will be dominated by a particular path or class of paths.  It is then sufficient to compare only the two dominant paths to determine which class of observers wins.

As an example, consider a toy landscape with two de Sitter vacua $A$ and $B$.  $B$ is connected to a terminal vacuum $T$ with negative cosmological constant.  Suppose also that only $B$ contains observers of any kind, and that ordinary observers are produced after $A$ decays to $B$. Then the dominant path to Boltzmann brains is represented as
\be I.C. \xrightarrow{P_B} B \xrightarrow{\Gamma_{BB,B}} BB \ [\xrightarrow{1'} B \xrightarrow{1'} T \xrightarrow{1} \mbox{crunch}]~. \ee
Here $1'$ above an arrow represents a branching ratio that is nearly unity.

Meanwhile, the dominant path to ordinary observers is
\be I.C. \xrightarrow{P_A} A\xrightarrow{1} OO \ [\xrightarrow{1'} B \xrightarrow{1'} T \xrightarrow{1} \mbox{crunch}]~. \ee  For completeness, we show a probable completion of the path in square brackets; this completion does not contribute to the amplitudes.

The total branching ratio for each type of observer is obtained by multiplying the probabilities for the corresponding dominant path.  One obtains $P_B\Gamma_{BB,B}$ for Boltzmann brains, and $P_A$ for ordinary observers.  Thus, an arrow of time is predicted if and only if $P_A/P_B>\Gamma_{BB,B}$.

\section{Two Toy Models}\label{toy}

\begin{figure*}[ht]
\centering
\includegraphics[scale=0.26]{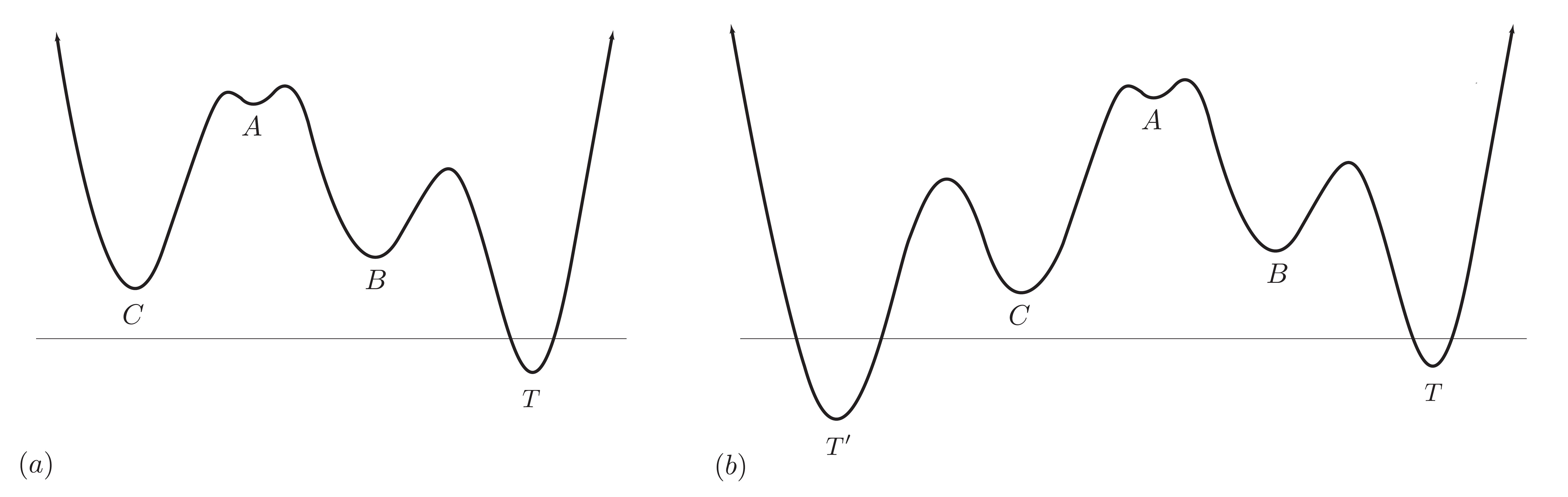}
\caption{(a) This toy landscape leads to an arrow of time even if initial conditions select the high entropy vacuum $C$.  This result persists if Hartle-Hawking initial conditions are chosen.  (b) With one extra terminal vacuum, an arrow of time is still predicted if initial conditions strictly select $C$.   However, Boltzmann brains dominate in the Hartle-Hawking state, which adds a small amplitude to start in $B$.}
\label{toymodel}
\end{figure*}

In this section, we consider the two toy landscapes shown in Fig.~\ref{toymodel}.  Both are one-dimensional, and only neighboring vacua are connected by decay. This is for simplicity, to illustrate the key physics, and general results will not depend on this assumption. The toy models differ from each other only through the number of terminal vacua.  Each theory predicts an arrow of time if initial conditions are entirely concentrated in the de~Sitter vacuum with highest entropy.   However, the two models may behave very differently under a tiny admixture of initial probability in other de~Sitter vacua.

\subsection{A Landscape With One Terminal}\label{sec-t1multi}

Consider the effective potential depicted in Fig.~\ref{toymodel}a. In this toy landscape, there are three de Sitter vacua $A, B$ and $C$, and one terminal vacuum $T$. It is assumed that ordinary observers are produced if $A$ decays to $B$; that both ordinary observers and Boltzmann brains can exist only in vacuum $B$; and that vacuum $B$ decays faster than it produces Boltzmann brains:
\begin{equation}
\Gamma_B > \Gamma_{BB,B}~.
\label{eq-assumption1}
\end{equation}
While simplistic, this toy model captures key features that are expected of the string landscape: for instance, there are large step sizes $|\Delta \Lambda| \gg S_{BB}^{-1}$ between vacua.

\subsubsection{Single-Vacuum Initial Conditions}
\label{sec-t1single}

Suppose now that initial conditions select the dominant vacuum $C$. From there the leading path to ordinary observer production is 
\be C \xrightarrow{1} A \xrightarrow{1} OO \ [\xrightarrow{1'} B \xrightarrow{1'} T \xrightarrow{1} \mbox{crunch}]~. \ee 
Naively, the second arrow has a branching ratio $\beta_{BA}$.  However, $C$ is a dead-end vacuum; if $A$ decays back to $C$, all that can happen is a return to $A$.  Strictly speaking one should sum over paths in which the entry to $B$ is preceded by any number of oscillations between $C$ and $A$, but the net effect is a branching ratio of $1$ from $A$ to $B$.  

The dominant path to Boltzmann brain production is
\be C\xrightarrow{1} A \xrightarrow{1} OO \xrightarrow{1'} B \xrightarrow{\Gamma_{BB,B}} BB \ [\xrightarrow{1'} B \xrightarrow{1'} T \xrightarrow{1} \mbox{crunch}]~. \ee

The Boltzmann path is suppressed by the extra factor $\Gamma_{BB} < e^{-S_{BB}}$, so ordinary observers win and an arrow of time is predicted.  (Note the importance of Eq.~(\ref{eq-assumption1}); with the opposite inequality, each path would contain of order $\Gamma_B$ additional Boltzmann brain production events before the vacuum $B$ decays to the terminal, and Boltzmann brains would win.)

\subsubsection{Multi-Vacuum Initial Conditions}
\label{eq-t1multi}
Now suppose each de~Sitter vacuum $A,B,C$ has initial probabilities $P_A, P_B$ and $P_C$ respectively. For each initial vacuum, we identify the dominant path to ordinary observers and the dominant path to $BB$s as follows:

\begin{align}
I.C. &\xrightarrow{P_C} C \xrightarrow{1} A \xrightarrow{1} OO \ [\xrightarrow{1'} B \xrightarrow{1'} T \xrightarrow{1} \mbox{crunch}]~.\\
I.C. &\xrightarrow{P_C} C \xrightarrow{1} A \xrightarrow{1} B \xrightarrow{\Gamma_{BB,B}} BB \ [\xrightarrow{1'} B \xrightarrow{1'} T \xrightarrow{1} \mbox{crunch}]~. \\
I.C. &\xrightarrow{P_A} A \xrightarrow{1} OO \ [\xrightarrow{1'} B \xrightarrow{1'} T \xrightarrow{1} \mbox{crunch}]~, \\
I.C. &\xrightarrow{P_A} A \xrightarrow{1} B \xrightarrow{\Gamma_{BB,B}} BB \ [\xrightarrow{1'} B \xrightarrow{1'} T \xrightarrow{1} \mbox{crunch}]~. \\
 I.C. &\xrightarrow{P_B} B \xrightarrow{\Gamma_{OO,B}} OO \ [\xrightarrow{1'} B \xrightarrow{1'} T \xrightarrow{1} \mbox{crunch}]~, \\
  I.C. &\xrightarrow{P_B} B \xrightarrow{\Gamma_{BB,B}} BB \ [\xrightarrow{1'} B \xrightarrow{1'} T \xrightarrow{1} \mbox{crunch}]~.
\end{align}

We see that with initial conditions purely in $C$ or $A$, ordinary observers would win.  But initial conditions in $B$ favor Boltzmann brains, since $\Gamma_{OO,B}\lll \Gamma_{BB,B}$. More generally, the outcome will, in principle, depend on the distribution of initial conditions. 

\paragraph {Ordinary observers dominate} if at least one of the following two conditions is satisfied:
\begin{align} 
\frac{P_B}{P_C} < \frac{1}{\Gamma_{BB,B}}~,\\
\frac{P_B}{P_A} <  \frac{1}{\Gamma_{BB,B}}~. 
\end{align}
Since $\Gamma_{BB,B} < e^{-S_{BB}}$, we find that a sufficient but not necessary condition is \be \frac{P_B}{P_C} < e^{S_{BB}} \label{1terminalOOcondition}~.\ee

\paragraph{Boltzmann brains dominate} if both of the following conditions hold:
\begin{align} 
\frac{P_B}{P_C} &> \frac{1}{\Gamma_{BB,B}}~,\label{BBcond1}\\ 
\frac{P_B}{P_A} &> \frac{1}{\Gamma_{BB,B}}~.\label{BBcond2} 
\end{align}

From Eq.~(\ref{1terminalOOcondition}), it is clear that ordinary observers will win for a large range of initial probabilities, including any case where initial conditions have the largest support in vacuum $C$. 

\subsection{A Landscape With Two Terminals}\label{sec-t2multi}

Now consider the landscape represented in Fig.~\ref{toymodel}b. The only difference is the extra terminal vacuum $T'$, which can be reached by a decay from $C$. This adds another feature expected to hold in the string landscape: every de~Sitter vacuum can decay to a vacuum with lower cosmological constant.  Consequently the branching ratio from $C$ to $A$ is no longer $1$.  The up-tunneling rate to $A$ contains a suppression factor $e^{-S_C}\ll e^{-S_{BB}}$.  Unless the decay rate to $T'$ is similarly suppressed, the branching ratio from $C$ to $A$ will be double-exponentially small.

\subsubsection{Single-Vacuum Initial Conditions}
\label{eq-t2single}

With initial conditions entirely in vacuum $C$, the extra terminal does not affect the conclusion that ordinary observers win.  The smaller branching ratio affects paths to $OO$s and $BB$s equally.  Paths from $C$ to $T'$ dominate overall but they are ``sterile'': they do not contribute to paths to either type of observers.

\subsubsection{Multi-Vacuum Initial Conditions}
\label{eq-t2multi}

However, the extra terminal has large implications for the case with more general initial conditions.  Suppose each de~Sitter vacuum $A,B,C$ has initial probabilities $P_A, P_B$ and $P_C$ respectively. For each initial vacuum, let us identify the dominant path to $OO$s and the dominant path to $BB$s:
\begin{align}
I.C. &\xrightarrow{P_C} C \xrightarrow{\beta_{AC}} A \xrightarrow{\beta_{BA}} OO \ [\xrightarrow{1'} B \xrightarrow{1'} T \xrightarrow{1} \mbox{crunch}]~.\\
I.C. &\xrightarrow{P_C} C \xrightarrow{\beta_{AC}} A \xrightarrow{\beta_{BA}} B \xrightarrow{\Gamma_{BB,B}}\ [\xrightarrow{1'} B \xrightarrow{1'} T \xrightarrow{1} \mbox{crunch}]~. \\
I.C. &\xrightarrow{P_A} A \xrightarrow{\beta_{BA}} OO \ [\xrightarrow{1'} B \xrightarrow{1'} T \xrightarrow{1} \mbox{crunch}]~, \\
I.C. &\xrightarrow{P_A} A \xrightarrow{\beta_{BA}} B \xrightarrow{\Gamma_{BB,B}} BB \ [\xrightarrow{1'} B \xrightarrow{1'} T \xrightarrow{1} \mbox{crunch}]~. \\
 I.C. &\xrightarrow{P_B} B \xrightarrow{\Gamma_{OO,B}} OO \ [\xrightarrow{1'} B \xrightarrow{1'} T \xrightarrow{1} \mbox{crunch}]~, \\
  I.C. &\xrightarrow{P_B} B \xrightarrow{\Gamma_{BB,B}} BB \ [\xrightarrow{1'} B \xrightarrow{1'} T \xrightarrow{1} \mbox{crunch}]~.
\end{align}

Again, with initial conditions purely in $C$ or $A$, $OO$s win.  But as before, initial conditions in $B$ favor Boltzmann brains since $\Gamma_{OO,B}\lll \Gamma_{BB,B}$. More generally, the outcome will depend on the distribution of initial conditions, and in this case we find this puts tighter restrictions on our initial conditions.

\paragraph {Ordinary observers dominate} if initial conditions sufficiently disfavor $B$ compared to at least one of $A$ and $C$.  That is, at least one of the following two conditions must be satisfied:
\begin{align} 
\frac{P_B}{P_C} &< \frac{\beta_{BA}\beta_{AC}}{\Gamma_{BB,B}}~,\label{2tOOcond1}\\
\frac{P_B}{P_A} &<  \frac{\beta_{BA}}{\Gamma_{BB,B}}~. \label{2tOOcond2}
\end{align}
These formulas have a very natural interpretation.  Ordinary observers dominate if the amplitude for starting in $B$ and then producing Boltzmann brains, $P_B \Gamma_{BB,B}$ is smaller than the amplitude for starting in some other vacuum and tunneling over to $B$, for example $P_C \beta_{BA}\beta_{AC} $. Note that in the latter case ordinary observers are automatically produced.

Since $\Gamma_{BB,B} < 1$, we find that a sufficient but not necessary condition is \be \frac{P_B}{P_C} < \beta_{BA}\beta_{AC}~. \label{OOcondition}\ee
As we will discuss in Sec.~\ref{toyeigen}, with dominant eigenvector initial conditions this condition is satisfied and hence $OO$s are predicted.

\paragraph{Boltzmann brains dominate} if both of the following conditions hold:
\begin{align} 
\frac{P_B}{P_C} &> \frac{\beta_{BA}\beta_{AC}}{\Gamma_{BB,B}}~,\label{BBcond1}\\ 
\frac{P_B}{P_A} &> \frac{\beta_{BA}}{\Gamma_{BB,B}}~.\label{BBcond2} 
\end{align}
As we will discuss in Sec.~\ref{HHtoy}, these conditions can both be satisfied with Hartle-Hawking initial conditions, leading to $BB$ domination.

\section{A Large Landscape}
\label{sec-large}

Now we generalize to the case of the large landscape, and derive sufficient conditions on the initial probability distribution for the existence or absence of an arrow of time.  We will make the following assumptions about the basic structure of the landscape:
\begin{itemize}
\item \emph{No tuning}. The low energy physics necessary for complex phenomena such as observers is fine-tuned, so that only a tiny fraction of vacua have observers of any type.
\item \emph{Large step size}. Vacuum transitions generically change the cosmological constant by a large amount $|\Delta \Lambda| \gg S_{BB}^{-1}$.
\item \emph{Not too large}. The effective number of vacua is less than $e^{S_{BB}}$. 
\item \emph{Not effectively one-dimensional}. For any two de Sitter vacua $i,j$ with $\Lambda_i, \Lambda_j < S_{BB}^{-1}$, there exists a semiclassical decay path from $i$ to $j$ that does not pass through any vacuum $k$ with $\Lambda_k < S_{BB}^{-1}$. 
\end{itemize}
These conditions are believed to be satisfied by the string landscape.  

We make one additional assumption, that all vacua decay faster than they produce Boltzmann brains:
\begin{equation}
\Gamma_{BB,i}<\Gamma_i~.
\label{eq-fastdecay}
\end{equation}
Whether this assumption holds in the landscape of string theory is not known though there is circumstantial evidence in its favor~\cite{Freivogel:2008wm}.

In the case of single-vacuum initial conditions, the above conditions are sufficient for the absence of Boltzmann brains.  We will review this argument in Sec.~\ref{sec-single}; it relies on the no-tuning assumption, which ensures (generically) that the single initial vacuum cannot produce Boltzmann brains at all.  In Sec.~\ref{sec-multi}, we consider the general case, allowing an admixture of initial vacua some of which may have a nonzero rate of producing Boltzmann brains.  The question is whether this spoils the prediction of an arrow of time.

\subsection{Single Initial Vacuum}
\label{sec-single}

Here we briefly review the case of single vacuum initial conditions in the string landscape; for further details, the original papers~\cite{Bousso:2011aa,Bousso:2008hz, DeSimone:2008if} should be consulted. We consider a slightly generalized setup where the initial vacuum is not necessarily the dominant vacuum, but the proof follows through exactly as before. Suppose the dominant path leading to both ordinary observers and Boltzmann brains starts in the de~Sitter vacuum $i$, with probability 1. There are two cases, depending on the size of the cosmological constant of the initial vacuum.

The first case is small initial vacuum energy: $\Lambda_{i} < S_{BB}^{-1}$. (In the toy model of the previous section, this corresponds to starting in vacuum $C$.) By the large-step-size property, vacua with observers of any type cannot be accessed directly from $i$.  Thus, all paths to observers of any kind begin with an up-tunneling from $i$ to a second vacuum $i_1$ with $\Lambda_{i_1} \gg S_{BB}^{-1}$, so
\be e^{S_{i_1}} \lll e^{S_{BB}}~. \ee
Detailed balance\footnote{Detailed balance is supported by explicit computation of Coleman de Lucia tunneling rates between de Sitter vacua, considered here. It does not apply to decays involving terminal vacua, which are irreversible at the semiclassical level.} relates the up-tunneling and down-tunneling rates between $i$ and $i_1$:
\be \Gamma_{i_1 i} e^{S_{i}} = \Gamma_{i i_1} e^{S_{i_1}}~. \label{eq-db} \ee
We also know that the decay time of $i_1$ cannot be larger than the recurrence time or faster than the Planck time, 
\be 1 > \Gamma_{i i_1} > e^{-S_{i_1}}~. \ee
Combining, we find
\be e^{S_{BB}} \ggg e^{S_{i_1}} > \Gamma_{i_1 i} e^{S_{i}} > 1 \ggg e^{-S_{BB}}~. \ee

The second vacuum $i_{1,OO}$ in the dominant path to ordinary observers need not be the same as the second vacuum $i_{1,BB}$ in the dominant path to Boltzmann brains.  However, the above inequalities imply that to an accuracy better than $e^{S_{BB}}$, the first up-tunneling suppresses paths to either Boltzmann brains or ordinary observers by the same amount:
\be e^{-S_{BB}} \lll \frac{\beta_{i_{1,OO}i}}{\beta_{i_{1,BB}i}} \lll e^{S_{BB}}~. \ee
This is sufficient accuracy to neglect the effects of the first up-tunneling, since we will find later that ordinary observers will be favored by a factor larger than $e^{S_{BB}}$. At this point the rest of the proof follows as in Case II, with $i_1$ now taken as the starting point. \\

The second case is large initial vacuum energy: $\Lambda_{i} > S_{BB}^{-1}$. (In the toy model of the previous section, this corresponds to starting in vacuum $A$.) By the assumed multi-dimensionality of the landscape, it is always possible to find a path through vacua that all have $\Lambda_{i_k} > S_{BB}^{-1}$ for all $k\geq 2$. The branching ratios through these paths will be less suppressed than $e^{-S_{BB}}$.  By the assumption that the landscape has fewer than $e^{S_{BB}}$ vacua, the sum over paths does not introduce any large factors for either type of observer. Let $e_{OO,i}$ and $e_{BB,i}$ be the total branching ratios for producing either $OO$s or $BB$s starting from $i$ (note that $OO$s and $BB$s do not in general need to be produced in the same vacuum). Then, including the first up-tunneling, we have
\begin{align} e_{OO, i} &> X_{OO} e^{-S_{BB}}~,\\
e_{BB, i} &< X_{BB} e^{-S_{BB}}~, \end{align}
where $X_{OO}$ and $X_{BB}$ differ by less than a factor of $e^{\pm S_{BB}}$. Thus to double-exponentially good accuracy,
\be e_{BB, i} < e_{OO, i}~, \label{singlevacuum} \ee
so $OO$s dominate over $BB$s given initial conditions with support only in $i$.

\subsection{General Initial Conditions}
\label{sec-multi}

Now we analyze the same large landscape, but with general initial conditions.  The probability for each class of observers can be computed by considering each initial vacuum with nonzero probability $P_i$, computing the expected number of $OO$s and $BB$s as in the single-initial-vacuum case, and then summing with weighting $P_i$.  

Naively, this means that $OO$s win. We showed in the previous subsection that they do so independently of the initial vacuum, so they should still win in a weighted average over initial vacua.  However, there is a key difference: we can no longer assume that all vacua with nonzero initial probability are unable to produce Boltzmann brains.  This contributes a new term to the expected number of Boltzmann brains, $\sum_j P_j \Gamma_{BB,j}$.  Therefore, ordinary observers win if and only if
\be \sum_j P_j \Gamma_{BB,j} < \sum_i P_i e_{OO, i}~, \label{condition}\ee
where each sum runs over all de Sitter vacua.

Because the landscape has fewer than $e^{S_{BB}}$ vacua, we expect that each sum is ``dominated'' by one path, in the following extremely weak sense: one can find a path such that dropping all other paths decreases the sum by less than a factor of $e^{S_{BB}}$.  This will be useful in some arguments below.

Note that the left hand side is always less than $e^{-S_{BB}}$, so ordinary observers win if the right hand side is greater than this double-exponentially small quantity.  For some theories of initial conditions, this is obviously the case: for example, if the probability is distributed evenly over all de~Sitter vacua, or with the tunneling wavefunction~\cite{Lin84b, PhysRevD.27.2848}.  In these cases, an arrow of time is predicted.  This is not surprising, since such initial conditions select for low initial entropy in any case.  What made the result of Ref.~\cite{Bousso:2011aa} interesting is that it was possible to obtain an arrow of time even after starting the universe in a state of arbitrarily large entropy.  

Thus, we will focus here on theories of initial conditions which select (or might select) for such initial states, but which give nonzero probability to several different initial vacua.  We will examine whether an arrow of time is still predicted when more than one initial vacuum is taken into account, i.e., whether Eq.~(\ref{condition}) is satisfied or violated.  In Sec.~\ref{HH} we  consider Hartle-Hawking initial conditions and find that they predict the absence of an arrow of time when the flow towards terminals is strong enough.  In Sec.~\ref{dominanteigenvector} we consider the initial conditions picked out by the global dual to the causal patch measure and closely related measures; we find that an arrow of time is predicted.

\section{Hartle-Hawking Initial Conditions}
\label{HH}
We now consider several specific choices of initial conditions and their implications for the existence of an arrow of time. In this section, we first analyze the case of Hartle-Hawking initial conditions. We will assume throughout the condition necessary for an arrow of time in the single vacuum case, i.e., that all vacua decay faster than they produce Boltzmann brains.  Hence the branching ratio to Boltzmann brains is small in all vacua:
\begin{equation}
\beta_{BB,j}=\Gamma_{BB,j}/\Gamma_j\ll 1~.
\end{equation}

\subsection{Hartle-Hawking Proposal}
In the Hartle-Hawking no-boundary proposal \cite{PhysRevD.28.2960}, the wave function of the universe, $\Psi[h_{ij},\ldots]$, is given by a path integral over all compact Euclidean four-manifolds whose only boundary is a given three-dimensional spacelike surface with metric $h_{ij}$. It would be nice to obtain probabilities directly in this framework, by computing the amplitude for specified field configurations, perhaps subject to additional conditions such as the presence of observers and some or all of their previous observations.  However, the measure problem cannot be circumvented so easily.  The data that can be conditioned on can only include what is available in the observer's past light-cone.  But the number of possible different quantum states in a past light-cone is finite~\cite{Bousso:1999xy,Bousso:1999cb,Bousso:2000nf,Bousso:2010pm}. There will be infinitely many saddlepoint geometries that contain the specified patch, and the sum over saddlepoints diverges.  Therefore, we will apply the Hartle-Hawking prescription only to obtain a probability distribution at an initial time.  We will use the causal patch measure or its close relatives to obtain well-defined probabilities in the resulting ensemble of semiclassical geometries.

The probability to start in a de~Sitter vacuum $i$, with cosmological constant $\Lambda_i$, is set by the action of the corresponding Euclidean de~Sitter instanton. It is proportional to the number of quantum states associated with empty de~Sitter space~\cite{Gibbons:1977mu}. In this section, we will find it convenient to work with unnormalized probabilities, i.e., we set
\begin{equation}
P_i=e^{S_i}=\exp(3\pi/\Lambda_i)~.
\end{equation}
Since we compute a relative probability for ordinary observers vs.\ Boltzmann brains, the overall normalization drops out in any case.  The probability to start in a state with $\Lambda\leq 0$ is assumed to vanish.

As a theory of initial conditions, the Hartle-Hawking proposal is problematic because as we will see, it exponentially favors initial conditions with large entropy.  (See Ref.~\cite{Page:2006hr} for a detailed discussion and further references.)  In the toy model of Sec.~\ref{sec-t1multi}, the proposal is nevertheless in accord with observation~\cite{Bousso:2011aa}.  However, we will show that the addition of another terminal vacuum destroys the prediction of an arrow of time if the initial vacuum can decay into it fast enough.  We will also examine more generally the conditions on a large landscape under which the Hartle-Hawking proposal is viable.  

\subsection{Toy Model}\label{HHtoy}
Hartle-Hawking initial conditions can lead to $BB$ domination in the toy model with three vacua and an extra terminal depicted in Fig.~\ref{toymodel}, as we will now show. We may neglect paths that start in $A$.

Boltzmann brains have unnormalized probability
\be P_{BB}=e^{S_B} \beta_{BB,B}= \frac{e^{S_B} \Gamma_{BB,B}}{\Gamma_B}= \frac{\mathcal{N}_{BB,B}}{\Gamma_B}~, \ee
where we have introduced the quantity
\begin{equation}
\mathcal{N}_{BB,B}\equiv e^{S_B} \Gamma_{BB,B}~.
\end{equation}
Since $e^{S_B}$ is the total number of quantum states associated with de~Sitter space, $\mathcal{N}_{BB,B}$ can be interpreted as the number of states that contain Boltzmann brains.  This number is always greater than $e^{S_{BB}}$, where $S_{BB}$ is the minimum coarse grained entropy of a Boltzmann brain.  But it may be much greater, since all other systems, and particularly the horizon, contribute to and typically dominate the entropy.

By detailed balance,
\be \Gamma_{AC} e^{S_C} = \Gamma_{CA} e^{S_A}~, \ee
so the unnormalized probability for ordinary observers is
\be P_{OO}= \frac{\beta_{BA}\Gamma_{CA}e^{S_A}}{\Gamma_{T'C}}~,\label{HHtoyOO}\ee
where we have assumed that $\Gamma_{T'C} \gg \Gamma_{AC}$.  Boltzmann brains win if and only if
\begin{equation}
\Gamma_{T'C}>\frac{\Gamma_B \beta_{BA}\Gamma_{CA}e^{S_A}}{{\cal N}_{BB,B}}~.
\end{equation}
Since $\Lambda_A > S_{BB}^{-1}$, the quantitities $\beta_{BA}$, $\Gamma_{CA}$, and $e^{-S_A}$ are all large compared to $e^{-S_{BB}}$, whereas ${\cal N}_{BB,B}^{-1}<e^{-S_{BB}}$.  By double-exponential arithmetic, Boltzmann brains win if and only if
\begin{equation}
\Gamma_{T'C}>\frac{\Gamma_B}{{\cal N}_{BB,B}}~. 
\label{eq-gtc}
\end{equation}
The right hand side is very small, so for a large range of parameters, the Hartle-Hawking proposal will not predict an arrow of time in this model.   However, note that Eq.~(\ref{eq-gtc}) does not involve the factor $e^{-S_C}$ associated with up-tunneling from the vacuum $C$.  Thus, it is not necessary to interpose a de~Sitter vacuum between $C$ and $T'$ to keep the  Hartle-Hawking viable.  It suffices to make the down-tunneling rate from $C$ to $T'$ smaller than the down-tunneling rate of $B$ (which could be quite large), divided by the number of Boltzmann states in vacuum $B$.\looseness=-1

\subsection{Large Landscape}

Our analysis of the toy model extends easily to the case of the large landscape considered in Sec.~\ref{sec-large}. With our choice of normalization, the amplitude for Boltzmann brains is closely related to the number of ``Boltzmann states," summed over all vacua:
\begin{equation}
P_{BB}=\sum_j e^{S_j} \beta_{BB,j} = \sum_j \frac{{\cal N}_{BB,j}}{\Gamma_j} \equiv \mathcal{N}~.
\label{eq-hhbb}
\end{equation}
Here ${\cal N}_{BB,j}= e^{S_j} \Gamma_{BB,j}$ is the number of quantum states, in the de~Sitter vacuum $j$, that contain at least one Boltzmann brain. This number is greater than $e^{S_{BB}}$, where the exponentially large number $S_{BB}$ is the coarse-grained entropy of a minimal Boltzmann brain.  It may be much greater since horizon entropy and the entropy of all other matter in the patch contributes to ${\cal N}_{BB,B}$.

The general amplitude for ordinary observers is given by 
\begin{equation}
P_{OO}=\sum_i e^{S_i} e_{OO,i}~.
\end{equation}
We can restrict the sum to vacua with $\Lambda_i\lesssim (\log {\cal N})^{-1} < S_{BB}^{-1}$.  Vacua with larger cosmological constant have $P_i<{\cal N}$; they are too suppressed by the Hartle-Hawking initial conditions to be able to compete with Eq.~(\ref{eq-hhbb}).  By arguments analogous to those in Sec.~\ref{sec-single}, we may further restrict the sum to a single initial vacuum and path, which dominates in the following weak sense: that dropping all other terms changes the sum by a factor less than ${\cal N}$.  Again up to factors double-exponentially smaller than ${\cal N}$, one then finds that the amplitude for ordinary observers can be estimated from the initial portion of this dominant path, keeping only the initial probability and the branching ratio for the first up-tunneling to the vacuum $i_1$:
\begin{equation}
P_{OO} \approx e^{S_i} e_{OO,i} \approx \beta_{i_1,i} e^{S_i} = \frac{\Gamma_{ii_1}e^{S_{i_1}}}{\Gamma_i}~.
\end{equation}
By the large step size property, $\Lambda_{i_1} > S_{BB}^{-1}$. Thus $e^{S_{i_1}}$ is larger than one and smaller than ${\cal N}$, since $1<e^{S_{i_1}} < e^{S_{BB}} < \mathcal{N}$. Similarly $\Gamma_{ii_1}$ is smaller than one and larger than ${\cal N}^{-1}$, since $1>\Gamma_{ii_1}>e^{-S_{i_1}} > \mathcal{N}^{-1}$.  By double exponential arithmetic, it follows that ordinary observers win if and only if
\begin{equation}
\Gamma_i < P_{BB}^{-1}~.
\label{eq-hhcond}
\end{equation}
If the sum in Eq.~(\ref{eq-hhbb}) is dominated by a single term $j$ (which is plausible), the necessary and sufficient condition for an arrow of time becomes
\begin{equation}
\Gamma_i < \frac{\Gamma_j}{{\cal N}_{BB,j}}~.
\label{eq-hhcond2}
\end{equation}
That is, the Hartle-Hawking proposal remains viable in a large landscape, if and only if ordinary observers are mainly produced along a path starting in a de~Sitter vacuum $i$ whose lifetime exceeds the lifetime of the most Boltzmann-friendly vacuum $j$ by a factor of the total number of Boltzmann states in $j$.

It is not known whether or not the string theory landscape satisfies Eq.~(\ref{eq-hhcond}).  Note that it is possible for a landscape to satisfy this condition, and to simultaneously satisfy the usual assumption that all vacua decay faster than they produce Boltzmann brains.

\section{Dominant Eigenvector Initial Conditions}
\label{dominanteigenvector}

Several of the most attractive measure proposals are related by global-local duality~\cite{Bousso:2012cv}. For instance, the scale-factor, light-cone time and CAH+ cutoffs are dual to the local ``fat geodesic'', causal patch, and Hubbletube cutoffs, respectively~\cite{Bousso:2009mw, Bousso:2008hz, Bousso:2012cv}. This duality holds for a very specific choice of initial conditions for the local measures, with probabilities given by the dominant eigenvector of the rate equation for eternal inflation. 

We will review the rate equation, the dominant eigenvector, and a method for computing the dominant eigenvector.   We will then argue that $OO$s will always win with these initial conditions.\looseness=-1

\subsection{Rate Equation}

Eternal inflation is the process by which a landscape of vacua is populated through exponential expansion and vacuum decay.  Consider a family of geodesics orthogonal to some fiducial hypersurface.  Let $f_j(t)$ be the fraction of the comoving volume occupied by vacuum $j$ at time $t$.  The rate equation describing the volume distribution is~\cite{Garriga:1997ef, Garriga:2005av}
\be \frac{df_j}{dt} = \sum_i (-\kappa_{ij} f_j + \kappa_{ji} f_i)~. \ee
This equation is appropriate to the scale factor time cutoff, which defines initial conditions for the fat geodesic by duality.  Essentially the same rate equation holds for the light-cone time cutoff, which is dual to the causal patch~\cite{Bousso:2012cv}.

We are interested only in the asymptotic distribution of de~Sitter vacua, which is governed by the matrix equation
\be \frac{df_i}{dt} = \sum_j R_{ij} f_j~, \ee
where
\be R_{ij} = \kappa_{ij} - \kappa_i \delta_{ij}~, \ee
and indices $i,j$ are now restricted to run over vacua with positive cosmological constant.
The total decay rates $\kappa_i$ will in general contain some contributions from decays to terminals. 

At late times, generic solutions to this equation evolve to an attractor distribution,
\be f_i(t) = s_i e^{-qt} + ... ~,\ee 
given by the dominant eigenvector $s_i$ of the transition matrix $R_{ij}$, that is, the eigenvector with the eigenvalue of smallest magnitude, $-q$~\cite{Garriga:2005av}:
\be \sum_j R_{ij} s_j = -q s_i~. \label{eigeneqn}\ee
Generically, the eigenvector will be dominated by the longest lived de Sitter vacuum, $*$: 
\begin{equation}
s_i\approx \delta_{i*}~.
\label{eq-sdelta}
\end{equation}
In previous work this approximation was assumed in the analysis of the Boltzmann brain problem.  Here we will go beyond the zeroth-order approximation and show that this does not change the conclusion.

\subsection{Toy Model}\label{toyeigen}

Let us revisit  the toy model depicted in Fig.~\ref{toymodel}b, with three de~Sitter vacua and two terminal vacua. We assume that $C$ is the longest-lived de~Sitter vacuum. We will derive the initial probability distribution by estimating the corrections to Eq.~(\ref{eq-sdelta}). 

The transition matrix is
\be R = 
\left(\begin{array}{ccc}
-\kappa_C & 0 & \kappa_{CA}\\
0 & -\kappa_B & \kappa_{BA}\\
\kappa_C \epsilon  & \kappa_{AB} & -\kappa_A \end{array}\right)~. \label{toymatrix}\ee
We assume that 
\begin{equation}
\epsilon\equiv\frac{\kappa_{AC}}{\kappa_C}\ll 1~.
\end{equation}
This is natural since tunneling from $C$ to $A$ is ``up-tunneling''; it increases the vacuum energy and decreases the entropy.   Note that $\kappa_A = \kappa_{BA} +\kappa_{CA}$. Expanding to first order in $\epsilon$, the eigenvalue $-q$ is written as
\be q = q^{(0)} + \epsilon q^{(1)}+...~, \ee
and the eigenvector is
\begin{align}
s = \left(\begin{array}{c}
1\\
s_B^{(0)} + \epsilon s_B^{(1)}+ ...\\
s_A^{(0)} + \epsilon s_A^{(1)}+ ...
\end{array}\right)~. \label{expandedeigenvector}
\end{align}
We have chosen a normalization in which the longest-lived vacuum has unit weight, for simplicity.  At leading order in $\epsilon$, the subleading entries will still be correctly normalized.

The eigenvalues are the roots of the characteristic polynomial, a cubic equation in $q$:
\begin{align} 
\mbox{Det}&(R+qI) = \epsilon \kappa_{C}\kappa_{CA}(\kappa_B-q)\nonumber \\
&-(\kappa_C-q)\left[-\kappa_{AB}\kappa_{BA}+(\kappa_B-q)(\kappa_A-q)\right]~.
\end{align}
At zeroth order in $\epsilon$, the correct choice is the smallest-magnitude root, $q^{(0)} = \kappa_C$. The first order correction is\looseness=-1
\be \epsilon q^{(1)} = -\frac{(\kappa_B-\kappa_C)\kappa_{CA}\kappa_{AC}}{(\kappa_B-\kappa_C)(\kappa_A-\kappa_C)-\kappa_{AB}\kappa_{BA}}~. \ee

Substituting Eq.~(\ref{expandedeigenvector}) into the rate equation Eq.~(\ref{eigeneqn}) with the matrix Eq.~(\ref{toymatrix}), and using our results for $q^{(0)}$ and $q^{(1)}$, we recover Eq.~(\ref{eq-sdelta}) at zeroth order:
\begin{align}
s_B^{(0)} &= 0~,\label{sB0}\\
s_A^{(0)} &= 0~.
\end{align}
At first order, we find
\begin{align}
\epsilon s_B^{(1)} &= \frac{\kappa_{BA}\kappa_{AC}}{(\kappa_B-\kappa_C)(\kappa_{A}-\kappa_C)-\kappa_{AB}\kappa_{BA}}~,\label{sB1}\\
\epsilon s_A^{(1)} &= \frac{(\kappa_B-\kappa_C)\kappa_{AC}}{(\kappa_B-\kappa_C)(\kappa_{A}-\kappa_C)-\kappa_{AB}\kappa_{BA}}~. 
\end{align}
As decay rates can be assumed exponentially small, and as $C$ is the longest-lived vacuum by assumption, we may set
\begin{equation}
\kappa_A-\kappa_C\approx \kappa_A~,~~~\kappa_B-\kappa_C\approx \kappa_B~.
\end{equation}
Thus, to first order in the up-tunneling branching ratio $\epsilon=\beta_{AC}$, we find that the initial probability for vacuum $B$ is given by
\be P_B\approx s_B \approx \frac{\kappa_{BA}\kappa_{AC}}{\kappa_B\kappa_{A}-\kappa_{AB}\kappa_{BA}} + O(\epsilon^2)~.\label{sB}\ee
Recall from Eq.~(\ref{2tOOcond1}) that a sufficient condition for ordinary observers to win in this toy model is
\be \frac{P_B}{P_C} < \frac{\beta_{BA}\beta_{AC}}{\Gamma_{BB,B}}~.\ee
With $P_C\approx 1$ and Eq.~(\ref{sB}), this condition reduces to
\be \beta_{AB}\beta_{BA}+\frac{\Gamma_{BB,B}}{\kappa_B}\kappa_C<1~, \label{conditiont2}\ee
This condition is easily satisfied, since Eq.~(\ref{eq-assumption1}) implies that both terms on the left hand side are double-exponentially small.  Recall that Eq.~(\ref{eq-assumption1}) [or more generally Eq.~(\ref{eq-fastdecay})] must be assumed; if it did not hold, Boltzmann brains would dominate already at zeroth order.

To see this in detail, we recall that $\Gamma_{BB,B}$ is double-exponentially small, whereas the powers of $\Lambda_B$ by which $\kappa_B$ differs from $\Gamma_B$ can be assumed to be at most exponentially small.  Thus, the second term in Eq.~(\ref{conditiont2}) can be approximated as $\frac{\Gamma_{BB,B}}{\Gamma_B}\kappa_C\ll\kappa_C\ll 1$.  Note also that vacuum $A$ must have more free energy than ordinary observers in vacuum $B$, which in turn have more free energy than Boltzmann brains in vacuum $B$.  This implies that $BB$ production in $B$ is enhanced compared to up-tunneling: $\Gamma_{AB}\ll \Gamma_{BB,B}$.  Together with Eq.~(\ref{eq-assumption1}), this implies $\beta_{AB}\ll \Gamma_{BB,B}\ll 1$.  Since $\beta_{BA}<1$, we conclude that the first term on the left hand side of Eq.~(\ref{conditiont2}), too, is very small.

Thus if ordinary observers win with initial conditions entirely in the longest-lived vacuum $C$, then they will still win when initial conditions are refined to reflect the dominant eigenvector at first order in the up-tunneling branching ratio from $C$, including the support in vacua other than $C$. Furthermore, they win easily: The probability to start in $B$ has a structure similar to the branching ratio along paths from the dominant vacuum $C$ to $B$. For $OO$s to win this must only be bounded by that exact branching ratio multiplied by the huge factor $\Gamma_{BB,B}^{-1}$. By comparison, recall that a sufficient condition Eq.~\ref{eq-gtc} for \emph{no} arrow in the same toy model with Hartle-Hawking initial conditions was also easily satisfied when the decay from $C$ to the terminal $T'$ was large compared to the double exponentially suppressed quantity $\Gamma_B \mathcal{N}_{BB,B}^{-1}$. 

\subsection{Large Landscape}

Consider the general landscape of Sec.~\ref{sec-large}, subject to the assumptions stated there.  These assumptions ensure that ordinary observers win with dominant eigenvector initial conditions, in the approximation that initial conditions have support entirely in the longest-lived de~Sitter vacuum.  We argue that this conclusion survives corrections that take into account that the dominant eigenvector has small support in other vacua as well.

In a terminal landscape, up-tunnelings will generally be suppressed compared to down-tunnelings. Thus it is appropriate to consider a general method of perturbation theory in up-tunneling branching ratios, which is discussed in detail in App.~\ref{pt}. At leading order, the correction to single-vacuum initial conditions is given by
\be s_i^{(n_0)}  = \sum_{p} \ \ \sideset{}{^{(n_0)}}\sum_{i_1,...,i_{p-1}} \frac{\kappa_{ii_{p-1}}}{D_i-D_*}\cdots \frac{\kappa_{i_1*}}{D_{i_1}-D_*}~. \ee
The superscript $(n_0)$ indicates summation over paths with exactly $n_0$ up-tunnelings and an arbitrary number of down-tunnelings. Approximating down-tunneling rates by total decay rates, and neglecting $\kappa_*$ compared to $\kappa_{i}$ for all $i$, we obtain
\be s_i^{(n_0)} = \sum_p \ \ \sideset{}{^{(n_0)}}\sum_{i_1,...,i_{p-1}} \frac{\kappa_*}{\kappa_i}\beta_{ii_{p-1}}\cdots\beta_{i_1 *}~.\label{leadingorder}\ee
Notice that this result involves precisely the product of branching ratios that determine $e_{i*}$ at the first nonzero order in the number of up-tunnelings.  Assuming both $e_i$ and $s_i$ are well approximated at this order, we may conclude that
\be s_i \approx \frac{\kappa_*}{\kappa_i} e_{i*}~. \label{eigenbound}\ee
Thus, the probability to start in vacuum $i$, $P_i\approx s_i$, is much smaller than the amplitude to decay to $i$ from $*$, $e_{i*}$. Multiplying each by the production rate of Boltzmann brains in vacuum $i$, the former yields the correction we seek while the latter, by the results of the previous section, is the zeroth order amplitude for Boltzmann brains.  Since $BB$s lose at zeroth order, this shows that the corrections cannot change the outcome.

To see this in detail, recall that we have already established from Eq.~\ref{singlevacuum} that
\begin{equation}
e_{BB,*} < e_{OO,*}~.
\label{eq-sv}
\end{equation}
This inequality implies the dominance of ordinary observers at zeroth order, i.e., with initial conditions purely in $*$.  Moreover, by Eqs.~(\ref{expectedn}), (\ref{newbranchingratios}), and~(\ref{eq-fastdecay}), the number of Boltzmann brains produced along paths that start in the $*$ vacuum is very small:
\be e_{BB,*} = \sum_j \frac{\kappa_{BB, j}}{\kappa_j} e_{j*}~. \label{eq-jkl} \ee
We will now combine these results with Eq.~(\ref{eigenbound}) to show that ordinary observers will still win with initial conditions given by the full dominant eigenvector.  

Technically, we must demonstrate that Eq.~(\ref{condition}) is satisfied.  Since $s_i$ is small for $i\neq *$, we can set $P_i\approx s_i$ and use Eq.~(\ref{eigenbound}) to bound the left hand side of Eq.~(\ref{condition}):
\begin{equation}
\sum_{j} P_j \Gamma_{BB, j}  \approx 
\kappa_* \sum_{j} \frac{\Gamma_{BB,j}}{\kappa_j} e_{j*}< \kappa_* e_{BB,*}~.
\end{equation}
The sum is taken over all $j$ since $\Gamma_{BB,*}=0$, so the contribution for $j=*$ vanishes. Also, in the second step we have used Eq.~(\ref{eq-jkl}) and the fact that $\Gamma_{BB,j}<\kappa_{BB,j}$ in Planck units. With Eq.~(\ref{eq-sv}), it follows that
\be \sum_{j} P_j \Gamma_{BB, j}  <\kappa_* e_{OO,*} \ll e_{OO,*} \approx  \sum_i P_i e_{OO, i}~, \ee
so that Eq.~(\ref{condition}) is indeed satisfied.  Ordinary observers win.

\bigskip
{\bf Acknowledgements}: We are grateful to D.~Page for alerting us to the importance of studying more general initial conditions.  We would also like to thank A.~Dahlen, S.~Leichenauer, V.~Rosenhaus and M.~P.~Salem for helpful discussions, and K.~Olum and D.~Schwarz-Perlov for correspondence. This work was supported by the Berkeley Center for Theoretical Physics, by the National Science Foundation (award numbers 0855653 and 0756174), by fqxi grant RFP3-1004, and by the U.S. Department of Energy under Contract DE-AC02-05CH11231. The work of CZ is supported by an NSF Graduate Fellowship. 

\appendix
\section{Perturbation theory}\label{pt}

The dominant eigenvector can be computed perturbatively in up-tunneling branching ratios~\cite{SchwartzPerlov:2006hi}. The rate equation matrix is separated into an upper triangular matrix $R^{(0)}$ consisting of down-tunnelings, and a lower triangular matrix containing up-tunneling rates, $R^{(1)}$:
\be R = R^{(0)} + R^{(1)}~. \ee
The entries on the diagonal separate into total down-tunnelings,
\be D_i = \sum_{j<i} \kappa_{ji}~, \label{D}\ee
which are included in $R^{(0)}$, and total up-tunnelings,
\be U_i = \sum_{j>i} \kappa_{ji}~, \ee
which are included in $R^{(1)}$. 

Due to the high suppression of up-tunneling, the eigenvector and corresponding eigenvalue can be computed perturbatively in $R^{(1)}$:\footnote{The perturbative approach breaks down if up- and down-tunneling branching ratios become comparable.  In a realistic landscape, this might be expected only for vacua near the Planck scale. For the purposes of the present analysis, however, no generality is lost in considering all states with $\Lambda > S_{BB}^{-1}$ as a single metastable de Sitter vacuum.  Then up-tunneling can occur only from vacua with a very small cosmological constant and will be extremely suppressed.}
\begin{align}
(R^{(0)} &+ \lambda R^{(1)})(s^{(0)} +\lambda s^{(1)} +\lambda^2 s^{(2)} + ...) \nonumber \\
&= -(q^{(0)} + \lambda q^{(1)} + ...)(s^{(0)} +\lambda s^{(1)} + \lambda^2 s^{(2)} + ...)~. \label{perturbeqn}
\end{align}
Equating the coefficients at each order in $\lambda$ gives a collection of matrix equations that can be iteratively solved for the eigenvector at each order in perturbation theory,

\small
\begin{align}
R^{(0)} s^{(0)} &= -q^{(0)} s^{(0)}~,\label{s0eqn}\\
(R^{(0)} + q^{(0)}I) s^{(1)} &= -(R^{(1)} + q^{(1)} I)s^{(0)}~, \label{s1eqn}\\
(R^{(0)} + q^{(0)}I) s^{(2)} &= -q^{(2)}s^{(0)}-(R^{(1)} + q^{(1)} I)s^{(1)}~,\label{s2eqn}\\
& \ \vdots \nonumber \\
(R^{(0)} + q^{(0)}I) s^{(n)} &= -\sum_{k=0}^{n-2}q^{(n-k)}s^{(k)} -(R^{(1)} + q^{(1)} I)s^{(n-1)}~. \label{sneqn}
\end{align}
\normalsize

At zeroth order, we posit that the unperturbed solution is a delta function with support only in the longest-lived de Sitter vacuum, which we denote $*$,
\be s_i^{(0)} = \delta_{i*}~. \label{s0}\ee
This selects the smallest magnitude eigenvector. (Note that this will only satisfy the zeroth order equation if there are no down-tunnelings from the dominant vacuum to non-terminal vacua~\cite{SchwartzPerlov:2006hi}.)

The entry of $R^{(0)} + q^{(0)}I$ corresponding to the dominant vacuum has a zero eigenvalue, rendering the matrix noninvertible. This means the system of equations is linearly dependent and has an infinite number of solutions; given Eq.~(\ref{s0}), one can still satisfy the $n$th order equation by adding to a solution $s^{(n)}$ any constant times $s^{(0)}$.

We fix this ambiguity by requiring orthogonality,
\be s^{(0)}\cdot s^{(n)} = 0\label{constraint}\ee
for all $n>0$. A different choice of constraint would correspond to re-normalizing and re-arranging the perturbative series. Our choice normalizes the dominant vacuum entry of the eigenvector to one,
\be s_* = 1~,\ee
so the eigenvector will no longer be normalized to $1$ when higher order effects are taken into account.  However, the criteria for $OO$ or $BB$ domination depend only on ratios of probabilities, and are thus independent of normalization.\looseness=-1

In general, the leading-order correction to Eq.~(\ref{s0}) may arise beyond first order, if it requires $n_0\geq 1$ up-tunnelings to reach a given de~Sitter vacuum $i$ from the $*$ vacuum. Before solving for this leading-order correction, we first consider the general solutions order-by-order up to $n$th order.

\subsection{Zeroth order}
We have required that the zeroth order solution to the eigenvector is given by Eq.~(\ref{s0}),
\be s_i^{(0)} = \delta_{i*}~. \label{s0b}\ee
From Eq.~(\ref{s0eqn}) we find the corresponding correction to the eigenvalue,
\be q^{(0)} = D_*~. \label{q0}\ee

\subsection{First order}
To solve for the first order correction, we invert Eq.~(\ref{s1eqn}), which is easy since the matrix is upper triangular. Each non-dominant entry gives an equation that can be solved iteratively:
 \be s_i^{(1)} = \hat\beta_{i*} + \sum_{j>i}\hat\beta_{ij} s_j^{(1)}~, \label{iteration}\ee
 where
  \be \hat \beta_{ij} \equiv \frac{\kappa_{ij}}{D_i-D_*}~. \ee

We claim that the entries $s_i^{(1)}$ of the solution to this equation (except for the entry $*$, which is zero) are sums of products of the factors $\hat \beta$ along paths from $*$ to the corresponding vacuum, which begin with one up-tunneling and have no further up-tunnelings, and pass through only non-dominant vacua at intermediate steps:
\be s_i^{(1)} = \sum_p \ \ \sideset{}{^{(1)}}\sum_{i_1,...,i_{p-1} \neq *} \hat \beta_{ii_{p-1}}\cdots \hat \beta_{i_1*}~, \label{s1}\ee 
where the superscript $(1)$ indicates that we are summing over paths with only one up-tunneling that start with an up-tunneling. $p$ labels path length, so the sum over $p$ indicates that we are summing over paths of all length that are consistent with this requirement on up-tunnelings.

We can prove this using induction. First, for the vacuum $N$ with the largest cosmological constant, Eq.~(\ref{iteration}) gives
\be s_N^{(1)} = \hat \beta_{i*}~, \ee
which satisfies Eq.~(\ref{s1}).

Now suppose that Eq.~(\ref{s1}) is satisfied by all entries $i>k$, and assume that the $k$th entry does not correspond to $*$. Then by Eq.~(\ref{iteration}),
\begin{align} 
s_k^{(1)} &= \hat \beta_{k*} + \sum_{j>k}\hat \beta_{kj} \sum_p \ \ \sideset{}{^{(1)}}\sum_{i_1,...,i_{p-1}\neq *} \hat \beta_{ji_{p-1}}\cdots\hat \beta_{i_1 *}\nonumber \\
&= \sum_p \ \  \sideset{}{^{(1)}}\sum_{i_1,...,i_{p-1}\neq*} \hat \beta_{ki_{p-1}}\cdots\hat \beta_{i_1 *}~,
\end{align} 
so Eq.~(\ref{s1}) is also satisfied by the $k$th entry. 

The final case occurs when $k=*$. In this case we replace the row with the constraint Eq.~(\ref{constraint}), which gives $s_*^{(1)} = 0$. Then proceeding to lower values of the cosmological constant, a similar argument shows that the $(k-1)$th entry satisfies Eq.~(\ref{s1}). This proves our claim.

Finally, substituting our result back into Eq.~(\ref{s1eqn}), we can solve the row $*$ for the corresponding correction to the eigenvalue,
\be q^{(1)} = -\sum_j \kappa_{*j}s^{(1)}_j + U_*~. \label{q1}\ee 

\subsection{$n$th order}
For the $n$th order case with $n>1$, extra terms are generated that naively seem to deviate from this simple pattern. To see this, we first invert Eq.~(\ref{sneqn}) as before. In terms of the compact notation
\be \Delta D_i = D_i - D_*~, \indent \Delta U_i = U_i - U_*~,\ee 
this gives the following iterative equation,
\small
\be s_i^{(n)} = \sum_{j<i}\hat\beta_{ij}s_j^{(n-1)} + \sum_{j>i}\hat\beta_{ij}s_j^{(n)}
 - \frac{U_i}{\Delta D_i} s_i^{(n-1)} + \sum_{k=0}^{n-1} \frac{q^{(n-k)}}{\Delta D_i} s_i^{(k)}~.\label{niteration}\ee
 \normalsize
The third term comes from fact that for $n>1$ the vector $s^{(n-1)}$ has components in non-dominant vacua, which multiply contributing diagonal entries of the matrix $R^{(1)} - q^{(1)} I$.

It is possible to check that the last term generates paths that return through the dominant vacuum; hence it does not contribute to the leading order result and so we will discard it here. It suffices to consider only the following simplified version of the full iterative equation: 
\be s_i^{(n)} \approx \sum_{j<i}\hat\beta_{ij}s_j^{(n-1)} + \sum_{j>i}\hat\beta_{ij}s_j^{(n)}
 - \frac{\Delta U_i}{\Delta D_i } s_i^{(n-1)}~.\label{niterationsimplified}\ee
We have included the up-tunneling factor $U_*$ contained in $q^{(1)}$ but set all other contributions from the last term in Eq.~(\ref{niteration}) to zero.

First, we can check---using an iterative argument that is very similar to the one we gave at first order---that the first two terms in this equation generate in the entries $s_i^{(n)}$ all sums of products of the factors $\hat\beta$ along paths from $*$ to the corresponding vacuum, which begin with an up-tunneling and have a total of $n$ up-tunnelings, and pass through only non-dominant vacua at intermediate steps: 
\be s_i^{(n)} \supset \sum_p \ \ \sideset{}{^{(n)}}\sum_{i_1,...,i_{p-1}} \hat \beta_{ii_{p-1}}\cdots \hat \beta_{i_1*}~, \label{sn}\ee 
where as before the superscript $(n)$ indicates that we are summing over paths with only $n$ up-tunnelings that start with an up-tunneling.

The effect of the third term in Eq.~(\ref{niterationsimplified}) is to add any number of factors $-\Delta U_{i_j}/\Delta D_{i_j}$ at intermediate or final vacua $i_j$ during the iteration. These factors appear non-trivially in front of terms in the path sum where the number of up-tunnelings in the path does not saturate the order $n$ in perturbation theory. If the entries vanish up to some order $n_0$, the leading order, this first nonzero order will not contain these factors since they multiply paths that would have appeared at lower order if they existed. Thus the leading order in up-tunnelings result is\looseness=-1
\be s_i^{(n_0)}  = \sum_{p} \ \ \sideset{}{^{(n_0)}}\sum_{i_1,...,i_{p-1}} \frac{\kappa_{ii_{p-1}}}{D_i-D_*}\cdots \frac{\kappa_{i_1*}}{D_{i_1}-D_*}~, \label{leadingorder}\ee
a sum over all possible paths with exactly $n_0$ up-tunnelings, which in general can have an arbitrary number of down-tunnelings. (A similar but inequivalent expression appears in Ref.~\cite{Olum07yk}.)

\bibliography{ArrowICs}

\end{document}